\documentclass[smallextended]{svjour3}       
\smartqed  
\usepackage{graphicx}
\usepackage{color}

\begin{document}

\newcommand{\be}{\begin{equation}}
\newcommand{\ee}{\end{equation}}
\newcommand{\bdm}{\begin{displaymath}}
\newcommand{\edm}{\end{displaymath}}
\newcommand{\bea}{\begin{eqnarray}}
\newcommand{\eea}{\end{eqnarray}}
\newcommand{\tb}[1]{\textbf{\texttt{#1}}}
\newcommand{\rtb}[1]{\textcolor[rgb]{1.00,0.00,0.00}{\tb{#1}}}
\newcommand{\btb}[1]{\textcolor[rgb]{0.00,0.00,1.00}{\tb{#1}}}
\newcommand{\il}{~}
\newcommand{\cf}{\textit{cf.}~}
\newcommand{\ie}{\textit{i.e.}~}

\newcommand{\oz}[1]{ \textcolor{red}   {\texttt{\textbf{OZ: #1}}} }

\title{Von Zeipel's theorem for a magnetized circular
  flow around a compact object}
\titlerunning{Von Zeipel's theorem for a magnetized circular
  flow}

\author{O. Zanotti         \and
        D. Pugliese 
}

\institute{O. Zanotti \at
              Universit\`a di Trento, Laboratorio di Matematica Applicata, Via Messiano 77, I-38123 Trento, Italy \\
              \email{olindo.zanotti@unitn.it}           
           \and
             D. Pugliese \at
              Institute of Physics, Faculty of Philosophy \& Science,
  Silesian University in Opava,
 Bezru\v{c}ovo n\'{a}m\v{e}st\'{i} 13, CZ-74601 Opava, Czech Republic
}

\date{Received: date / Accepted: date}

\maketitle

\begin{abstract}
We analyze a class of physical properties, forming the content of the so-called von Zeipel theorem,
which characterizes stationary, axisymmetric, non-selfgravitating perfect
fluids in circular motion in the gravitational field of a compact object.
We consider the  extension of  the theorem to the
magnetohydrodynamic regime, under the assumption of an
infinitely conductive fluid,
both in the Newtonian and in the relativistic framework.
When the magnetic field is toroidal, the conditions required by the theorem
are equivalent to integrability conditions, as it is the
case for purely hydrodynamic flows.
When the magnetic field is poloidal, the analysis for the
relativistic regime is substantially different with respect to the Newtonian case and
additional constraints, in the form of PDEs,
must be imposed on the magnetic field in order to guarantee that the angular velocity $\Omega$ depends only on the specific angular momentum $\ell$.
In order to deduce such physical constraints, it is crucial to adopt special coordinates, which are adapted to the $\Omega={\rm const}$ surfaces.
The physical significance of these results is briefly discussed.
\end{abstract}

\section{Introduction}

In Newtonian hydrodynamics, the so-called
{\em von Zeipel theorem} \cite{vonZeipel1924} specifies the physical conditions
to which all stationary, axially symmetric, perfect fluids
in circular motion must obey in order for the angular velocity $\Omega$ to
depend only on the distance from the rotation axis.
Abramowicz formulated the relativistic
version\footnote{
The original version of the theorem by von Zeipel considers the role of a radiation field (see \cite{Claret2012} for a modern account), while
in this paper we have in mind the ``von Zeipel's theorem in General Relativity", called in this way for the first time presumably by Thorne \cite{Abramowicz71} and which has nothing to do with radiation effects.
}
of the von Zeipel theorem~\cite{Abramowicz71,Abramowicz1974}, highlighting
that the angular velocity $\Omega$ and the specific angular momentum $\ell$ have common iso-surfaces, with the topology of a cylinder,
if and only if the rotating fluid is barotropic.
This property was used soon after to
build equilibrium solutions of geometrically thick disks (tori) around
black holes~\cite{Fishbone76,Abramowicz78,Kozlowski1978}.
In fact, the von Zeipel theorem for
a purely hydrodynamic flow represents a set of integrability conditions for computing
the equilibrium solution.
Over the years, these analytic or semi-analytic
solutions, sometimes referred to as
``Polish doughnuts'', turned out to be very useful to
study various kinds
of fluid instabilities and potentially
detectable physical effects
around compact objects, either black
holes or neutron stars~
\cite{Abramowicz1980,Abramowicz83,Papaloizou84,Blaes1988,Abramowicz1998,Rezzolla_qpo_03b,Zanotti03,Blaes2006,Montero2010,Pugliese2013,Pugliese01112012,Slany2013}.

Though rather simplified,
geometrically thick discs are still attracting a lot of
interest in high energy astrophysics. From one side, they are currently adopted as
initial conditions in general relativistic
hydrodynamic and magnetohydrodynamic numerical simulations of
accretion flows~\cite{DeVilliers03,Fragile2009b,Barkov2011,Narayan2012,McKinney2012},
even in the presence of radiation
fields~\cite{Sadowski2013b,McKinney2013b,Zanotti2014}.
Moreover, an additional indication supporting their
astrophysical relevance is provided by the outcome of
fully relativistic numerical
simulations, which clearly show that
high-density tori are indeed produced after the merger of
neutron star
binaries collapsing onto a black hole~\cite{Rezzolla:2010}.

A substantial but necessary complication in the study of
geometrically thick discs is represented by magnetic fields,
as it is now generally accepted
that in accretion flows
magnetic fields are of fundamental
importance to account for the outward transport of angular
momentum induced
by the magnetorotational instability.
A  few years ago,
the Polish doughnut model was extended to flows with a toroidal magnetic
field~\cite{Komissarov2006a,Admek2013,Wielgus:2014nva} and later adopted in various astrophysical applications~\cite{Montero07,Hamersky2013}.
However, in these works it is not  discussed whether
and how the von Zeipel theorem can be rephrased when a magnetic field is present.
In this paper we fill
this gap by studying the von Zeipel theorem
for a magnetized circular
flow around a compact object,
 by considering, separately, the case of a toroidal and of a poloidal magnetic field. While for the former case we find that von Zeipel theorem still provides an integrability condition, which is met, for instance,  in
the model by \cite{Komissarov2006a}, for the latter
case the magnetic field needs to be constrained by two additional partial differential equations.
We emphasize that it is not our intention to compute an equilibrium model,
which would involve the solution of the Grad-Shafranov equation,
but rather to clarify
the physical conditions required by the von Zeipel theorem for a magnetized flow.
We also stress that in some circumstances, like for instance in the interior of neutron stars, the magnetic field has a twisted topology
(see \cite{Ciolfi2009,Ciolfi2014}), but in our analysis we have not considered this possibility, nor the issue of the various magnetic instabilities which may arise.

Since Newtonian
physics can still provide useful comparisons, we
first start from the Newtonian version of the von Zeipel
theorem.
Although our discussion is motivated by the astrophysical applications
mentioned so far, in what follows
we do not assume any particular form of the spacetime metric.
Our only assumptions will be those of
stationarity and axisymmetry for both gravity and matter, and of a perfect,
infinitely conductive non-selfgravitating
fluid in purely circular motion around a compact object.

We set the speed of light $c=1$,
the gravitational constant $G=1$,
and we adopt the
Lorentz-Heaviside notation for the electromagnetic
quantities, such that all $\sqrt{4\pi}$ factors disappear.
Greek indices run from $0$ to $3$,
Latin indices run from $1$ to $3$
and we use the Einstein
summation convention of repeated indices.

\section{Newtonian version}
\label{sec:2}

\label{newt_vs}
The Newtonian version of von Zeipel's theorem in the
absence of magnetic fields states
that the iso-density and iso-pressure surfaces within a
rotating object coincide if
and only if the
angular velocity is a function of the distance
from the rotation axis only \cite{Tassoul-1978:theory-of-rotating-stars}.
If a magnetic field is present, a weaker version of
this theorem can be proved.
To this scope, we consider a stationary and
axisymmetric system and a set of
cylindrical coordinates $(r,\phi,z)$.
The two relevant Euler equations can be written as
\begin{equation}
\label{euler}
\partial_i \Phi + \frac{1}{\rho}\partial_i p -\delta^r_i
\Omega^2 r = \frac{1}{\rho}((\vec\nabla\times\vec {B})\times
\vec{B})_i\,,  \hspace{1cm} i=r,z  \,,
\end{equation}
where $\Phi$ is the gravitational potential, $\rho$ is
the gas density, $p$ is the pressure, $\Omega$ is the
angular velocity of the rotating fluid
and $\vec{B}$ is the magnetic field.
To simplify our treatment, in the following we
separately consider the case of a purely toroidal and of
a purely poloidal magnetic field, while keeping equation
(\ref{euler}) as the reference equation.

\subsection{Toroidal magnetic field}
\label{tn}
When the magnetic field is toroidal, the components of
the poloidal vector on the right hand side of equation
(\ref{euler}) are given by
\begin{eqnarray}
((\vec\nabla\times \vec{B})\times \vec{B})_i&=&-\partial_i (r
B_\phi)\frac{B_\phi}{r}\,, \hspace{1cm} i=r,z  \,.
\end{eqnarray}
If  we take the partial derivative  $\partial_j$ of
(\ref{euler}) and then
multiply it  with
$e_{kji}$, all the symmetric terms in $i$ and $j$
disappear, and we are left with
\begin{equation}
\label{333}
e_{kji}\frac{\partial_j
\rho\partial_i p}{\rho^2} +
 e_{kji} \partial_j(\delta^r_i \Omega^2
r)
=
e_{kji}\partial_i(r
B_\phi)\partial_j\left(\frac{B_\phi}{r\rho}\right)\,.
\end{equation}
In the absence of a magnetic field, the right hand side
of the above equation
vanishes and we would have the statement of the
theorem as reported in \cite{Tassoul-1978:theory-of-rotating-stars}.
On the contrary, if the magnetic field is not zero, then
we can only provide a sufficient condition for the
theorem.
Namely, we
can say that if
\begin{enumerate}
\item the equation of state is barotropic, $p=p(\rho)$, and
\item (a) $B_\phi\sim r^{-1}$, or,  (b) $B_\phi\sim r \rho $, or
      (c) $B_\phi/r\rho=f(r B_\phi)$,
\end{enumerate}
then the angular velocity would only
depend on the distance from the rotation axis.

Note, for instance, that \cite{Okada1989}
proposed an equilibrium model for
a barotropic, magnetized and geometrically thick disc
around a central object by assuming that
the toroidal magnetic field scales like $B_\phi^2\sim
r^{2(\mu-1)}\rho^\mu$. This choice indeed satisfies von Zeipel's
theorem and it corresponds to constraint 2(c) above, with $f(r
B_\phi)= (r B_\phi)^{1-2/\mu}$.
It should be noted that the fulfillment of von
Zeipel's hypothesis not only guarantees that
$\Omega=\Omega(r)$ but it also provides the functional
dependencies among $B_\phi$, $\rho$ and $r$ that allows to
write the
Euler equation in a potential form. In fact, we essentially
imposed the vanishing of the curl of the right hand side
of equation (\ref{euler}), so we looked for those cases
when $((\vec\nabla\times\vec B)\times\vec B)/\rho$ can be
written as the gradient of a scalar.

\subsection{Poloidal magnetic field}
In cylindrical coordinates, a poloidal magnetic field
can be written as
\begin{equation}
\vec B =\frac{1}{r}\vec{\nabla} \Psi \times
\vec{e_\phi}=\left(-\frac{\partial_z\Psi}{r},0,\frac{\partial_r
\Psi}{r}\right) \ ,
\end{equation}
where $\vec{e_\phi}$ is the unit vector along the $\phi$
direction and
$\Psi$, that physically represents
the magnetic flux through a $z=z_0$, $r=r_0$ circle,
is the $\phi$-component of the electromagnetic vector
potential.\footnote{In non-relativistic studies, $\Psi$ is usually called the \emph{magnetic stream function} or \emph{magnetic flux function}.}
While in the case of toroidal magnetic fields
the Maxwell equations $\vec{\nabla}\cdot\vec{B}=0$ and
$\vec{\nabla}\times \vec{E}=0$ are automatically satisfied under the assumption of stationarity and axisymmetry,
when the magnetic field is poloidal they need to be
properly taken into account, and they lead
to  so called Ferraro's iso-rotation law, namely
$\vec{B}\cdot\vec{\nabla}\Omega=0$.
Hence, a necessary condition to have
$\Omega=\Omega(r)$ is that $B^r=0$, or, equivalently,
that $\Psi=\Psi(r)$.
As a result,
\begin{eqnarray}
(\vec\nabla\times\vec{B})\times \vec{B}&=&
(- B_z\partial_r B_z,0,0) \ ,
\end{eqnarray}
and by computing the curl of equation (\ref{euler})
we find
\begin{eqnarray}
\label{GG1}
2\Omega
r\vec{\nabla}\Omega\times
\vec{e}_r=
-\frac{\vec{\nabla}\rho}{\rho^2}\times\vec{\nabla}p
-\frac{B_z}{\rho^2}\partial_z\rho\,\partial_r
B_z\, \vec{e}_\phi \ .
\end{eqnarray}
From (\ref{GG1}), a weaker version of
von Zeipel's theorem
for a poloidal magnetic field can be deduced. Namely, if
\begin{enumerate}
\item the equation of state is barotropic, $p=p(\rho)$,
\item $B^r=0$, and
\item (a) $B_z={\rm const}$, or,  (b) $\rho=\rho(r)$\,,
\end{enumerate}
then $\Omega=\Omega(r)$.
We note, incidentally, that a Newtonian star
with a dipolar magnetic field will not satisfy this
version of von Zeipel's theorem, since $B^r\neq 0$.

\section{General relativistic version}
\label{gr_vs}
The general relativistic version of von Zeipel's theorem in the absence of
magnetic fields is due to \cite{Abramowicz71}
and it
states that in a stationary and
axisymmetric system the surfaces of constant angular
velocity $\Omega$ and
the surfaces of constant specific angular momentum $\ell$
coincide if and only if the rotating fluid is barotropic,
i.e. it has an equation of state $p=p(e)$, where $e$ is
the total energy density. Now we
consider how this theorem can be rephrased if magnetic
fields are present, again by distinguishing between the
two main magnetic field topologies.
Let us first recall the form of the  energy momentum tensor
for an magnetized flow with infinite conductivity, namely \cite{Anile_book}
\begin{equation}
T^{\alpha\beta}=(\omega+b^2) u^\alpha
u^\beta+(p + b^2/2) g^{\alpha\beta}- b^\alpha b^\beta \ ,
\end{equation}
where $u^\alpha$ are the components of the four velocity of
the fluid,  $b^\alpha$ are the components of the four vector
magnetic field, $b^2=b_\alpha b^\alpha$, $g_{\alpha\beta}$ are the coefficients of the
metric, while $\omega=e+p$ is
the enthalpy density.
If we adopt a system of coordinates $(t,x^1,x^2,\phi)$ where $t$ and
$\phi$ are the
coordinates associated to the temporal and axial Killing
vectors, respectively, then
the four velocity of the circular motion is
$u^\alpha=u^t(1,0,0,\Omega)$, where $\Omega=u^\phi/u^t$ is
the angular velocity.
Finally, the
specific angular momentum mentioned in the theorem is
$\ell=-u_\phi/u_t$, which is related to $\Omega$ by
\be
\Omega = -\frac{g_{t\phi} + g_{tt}\ell}{g_{\phi\phi} + g_{t\phi}\ell}\,.
\ee
The relativistic Euler equation can be written as
\cite{Bekenstein1978}
\begin{eqnarray}
\label{euler0}
\nonumber
&&(\omega+b^2)a_\alpha + \nabla_\alpha\left(p+\frac{b^2}{2}\right) +
  u_\alpha u^\pi\nabla_\pi(p+b^2) + u_\alpha b^2
  \nabla_\pi u^\pi - \nabla_\pi(b_\alpha
  b^\pi)=0 \ .
\end{eqnarray}
As a result of the symmetries,
$u^\alpha\nabla_\alpha(p+b^2)=\nabla_\alpha u^\alpha=0$.
Moreover, the four acceleration for the circular motion is \cite{Abramowicz71}
\begin{equation}
\label{acceleration}
a_\alpha = -\nabla_\alpha \ln u^t +
\frac{\ell}{1-\Omega\ell}\nabla_\alpha\Omega \,.
\end{equation}
Combining all this, the Euler equation (\ref{euler0}) becomes
\begin{equation}
\label{gr_euler}
-\nabla_\alpha \ln u^t +
u^t u_\phi \nabla_\alpha\Omega
= -\frac{\nabla_\alpha(p+b^2/2)}{\omega+b^2} +
\frac{\nabla_\pi(b^\pi b_\alpha)}{\omega+b^2}\,,
\end{equation}
where $u^t u_\phi=\ell/(1-\Omega \ell)$.
In ideal magnetohydrodynamics,
the fluid acceleration obeys a few additional contraction relations, namely  \cite{Anile_book}
\be
\label{eq:ab}
a_\alpha b^\alpha = \nabla_\alpha b^\alpha = -\frac{b^\alpha\partial_\alpha p}{\omega}\,,
\ee
that will be used below.
Finally, according to Ferraro's iso-rotation law valid
for stationary and axisymmetric systems,
the normals to the equipotentials of
$\Omega$ and of $\Psi$ are parallel,
hence $\Omega=\Omega(\Psi)$, (see \cite{Mason1977,Bekenstein1978,Markakis2011}).

\subsection{Toroidal magnetic field}
When the magnetic field is purely toroidal, it has been shown by \cite{Komissarov2006a}
that Eq.~(\ref{gr_euler}) can be
rewritten as
\begin{equation}
\label{kom}
- \nabla_\alpha \ln u^t +
u^t u_\phi
\nabla_\alpha\Omega +
  \frac{\nabla_\alpha p}{\omega} + \frac{\nabla_\alpha
  ({\cal L}b^2)}{2{\cal L}\omega}=0  \  ,
\end{equation}
where ${\cal L}=g_{t\phi}^2 - g_{tt}g_{\phi\phi}$. If we
now take the covariant derivative $\nabla_\beta$ of
(\ref{kom}) and then we multiply it by the completely
antisymmetric tensor $\epsilon^{\alpha\beta\gamma\delta}$,
all the symmetric terms in $(\alpha,\beta)$ vanish, and
we find
\begin{eqnarray}
\label{gr_euler_2}
\epsilon^{\alpha\beta\gamma\delta}\nabla_\alpha\Omega\nabla_\beta(u^t
u_\phi)
=&&\epsilon^{\alpha\beta\gamma\delta}\frac{\nabla_\alpha p
\nabla_\beta \omega}{\omega^2}-\epsilon^{\alpha\beta\gamma\delta}\nabla_\alpha({\cal L}
  b^2)\nabla_\beta\left(\frac{1}{2{\cal L}\omega}\right)\,.
\end{eqnarray}
In the absence of the magnetic field, the last term on the
right hand side of this equation would vanish, and we
would have the statement of von Zeipel's theorem as
provided in \cite{Abramowicz71}.
In the presence of a magnetic field, a weaker version of von
Zeipel's theorem
can  be formulated by saying that
$\Omega={\rm const}$ surfaces coincide with $\ell={\rm const}$ surfaces if
\begin{enumerate}
\item the equation of state is barotropic, i.e. $p=p(e)$, and
\item (a) ${\cal L}
b^2=const$, or (b) ${\cal L}\omega=const$, or (c) ${\cal
L}b^2=f({\cal L} \omega)$\,.
\end{enumerate}
Note that, for instance,
in the equilibrium model proposed by \cite{Komissarov2006a}
$b^2\sim {\cal L}^{\eta-1}\omega^\eta$, with $\eta$ a
real index, and this choice
does satisfy von Zeipel theorem as it corresponds
to constraint 2(c), with $f$ a power law, i.e. ${\cal L}
b^2\sim ({\cal L} \omega)^\eta$.
Also note that, as expected, the Newtonian limits of constraints 2(a),
2(b) and 2(c) just found span the same physical conditions provided
by their Newtonian analogs that we found in Sec.~\ref{tn}.

\subsection{Poloidal magnetic field}

In the relativistic
framework,
the treatment of the poloidal magnetic field is
significantly more involved than that of toroidal magnetic fields, and in what follows we use several results proved by
Bekenstein \& Oron \cite{Bekenstein1978,Bekenstein1979}.
We first exploit the freedom in the choice of the
coordinates
system to adopt coordinates
$x^1=z$ and $x^2=\chi$
such that $\chi$ is
constant along $\Psi={\rm const}$ surfaces,
i.e. $\Psi=\Psi(\chi)$,
while $z$ is
constant along the normals to those surfaces.
In the following, and just for notational convenience, we will refer to $(z,\chi)$ as to the B-O coordinates.
An example of such coordinates is reported in Appendix A for the
Schwarzschild metric.

We now recall a property highlighted by \cite{Oron2002},
who showed that a spacetime containing a purely toroidal
flow with a purely poloidal (or purely toroidal) magnetic
field is circular, meaning that in the chosen B-O coordinates the metric can be written
in ``diagonal plus one'' form  as
\begin{equation}
\label{metric}
ds^2=g_{tt}dt^2 + 2g_{t\phi}dt d\phi +
g_{\phi\phi}d\phi^2 + g_{zz}dz^2 + g_{\chi\chi}d\chi^2 \ .
\end{equation}
Moreover, from the definition of magnetic field in terms
of the electromagnetic tensor, namely
$b^\alpha=^*F^{\alpha\beta} u_\beta$, where
$^*F^{\alpha\beta}$ is the dual of  $F^{\alpha\beta}$,
using the fact that there is no meridional motion
($u^z=u^{\chi}=0$) and that $F_{\alpha\beta}u^\beta=0$ (because
of infinite conductivity) one finds (\cite{Bekenstein1979})
\begin{eqnarray}
b^t&=&b^\phi=0\,,\\
b^i&=&\frac{\epsilon^{ij}}{u^t\sqrt{-g}}\partial_j\Psi
\hspace{1cm} i,j=z,\chi  \ ,
\end{eqnarray}
where $\epsilon^{z\chi}=-\epsilon^{\chi z}=1$,
$\epsilon^{zz}=\epsilon^{\chi\chi}=0$ and where $g$ is the
determinant of the metric (\ref{metric}).
Therefore, the magnetic field lines lie on the equipotentials
of $\Psi$, namely
\be
\label{eq:bPsi}
b^\alpha\partial_\alpha \Psi=0\,,
\ee
which, due to Ferraro's iso-rotation law, also implies
\be
\label{eq:bOmega}
b^\alpha\partial_\alpha \Omega=0\,.
\ee
We therefore expect that the magnetic field lines lie on the surfaces of
constant magnetic potential $\Psi$ (magnetic surfaces), which coincide
with the surfaces of constant angular velocity $\Omega$. This
property prevents the generation of a toroidal component of the magnetic
field, even in the presence of differential
rotation.\footnote{If the poloidal magnetic field lines lie on the $\Psi={\rm const}$ surfaces, and if these surfaces coincide with the $\Omega={\rm const}$ surfaces, then the magnetic field lines will not ``feel" rotational effects.}
The choice of the coordinates $(z,\chi)$ as
described above turns
out to be rather useful, because
the poloidal magnetic field components are
\begin{eqnarray}
\label{eq:bz}
b^z&=&\frac{1}{u^t\sqrt{-g}}\partial_\chi \Psi\,,\\
\label{eq:bchi}
b^\chi&=&0\,,
\end{eqnarray}
while, from Eq.~(\ref{eq:bOmega}), it also follows that
\be
\partial_z\Omega=0 \Longrightarrow \Omega=\Omega(\chi)=\Omega(\Psi)\,.
\ee
{
We note that the coordinate $\chi$ resembles closely the cylindrical coordinate $r$ of the Newtonian case, for which we found
$B^r=0$ from Ferraro's iso-rotation law, in analogy to Eq.~(\ref{eq:bchi}).}
By combining Eq.~(\ref{eq:ab}) and Eq.~(\ref{eq:bchi}),
the fluid acceleration along the $z$-direction can be written as
 \be
\label{Eq:rev-Eq}
 a_z=-\partial_z\ln u^t=\frac{\nabla_\alpha b^\alpha}{b^z}=-\frac{\partial_zp}{\omega}\,.
 \ee
Focusing on the last equality of Eq.~(\ref{Eq:rev-Eq})
and writing explicitly the four-divergence of a
four-vector, we obtain
\be\label{Eq:for-bz}
\partial_z\ln \left(\frac{b\sqrt{-g}}{\sqrt{g_{zz}}}\right)=-\frac{\partial_z p}{\omega}\,,
\ee
where $b=(b^z b_z)^{1/2}$.
Equation\il(\ref{Eq:for-bz}) places a first condition on the
derivative $\partial_z b$, and can be used to write
$\partial_zb/b$ as function of the metric and of the
field. The interesting aspect of this equation is that it only involves derivatives along the coordinate $z$.
We emphasize that it has nothing to do with von Zeipel's theorem, and it is a consequence of the adopted B-O coordinates.

The Euler equation (\ref{gr_euler}) can also be written in terms of the new coordinates $(z,\chi)$. Before doing that, we write the magnetic terms on the right hand side of Eq.~(\ref{gr_euler}) as
\begin{eqnarray}\label{Eq:relation}
-\frac{1}{2}\nabla_\alpha b^2 + \nabla_\pi(b^\pi
 b_\alpha)=b^\pi(\partial_\pi b_\alpha - \partial_\alpha
 b_\pi) - b_\alpha b^z \partial_z\ln u^t \ ,
\end{eqnarray}
where we have again used Eq.~(\ref{eq:ab}),
with $a_z$ given by
Eq.~(\ref{Eq:rev-Eq}).
Hence,
we can now write (\ref{gr_euler}) in the two components $\chi$
and $z$ to obtain
\begin{eqnarray}
\label{term1}
-\partial_\chi \ln u^t +
u^t u_\phi \partial_\chi\Omega
&=& -\frac{\partial_\chi p}{\omega+b^2} -
\frac{b^z\partial_\chi b_z}{\omega+b^2} \\
\label{term2}
-\partial_z \ln u^t
&=& -\frac{\partial_z p}{\omega+b^2} -
\frac{b^2 \partial_z\ln u^t}{\omega + b^2} \ .
\end{eqnarray}
After taking the z-derivative of the first equation and the $\chi$-derivative of the second one, we find
\begin{eqnarray}
\label{terms1s}
-\partial_z\partial_\chi \ln u^t +
\partial_z(\mathcal{K}(\ell) \partial_\chi\Omega)
&=& \partial_z\left(-\frac{\partial_\chi p}{\omega+b^2}\right) -
\partial_z\left(\frac{b^z\partial_\chi b_z}{\omega+b^2} \right)\\
\label{terms2s}
-\partial_\chi\partial_z \ln u^t &=& -\partial_\chi\left(\frac{\partial_z p}{\omega+b^2}\right) -
\partial_\chi\left(\frac{b^2 \partial_z\ln u^t}{\omega + b^2} \right)\,,
\end{eqnarray}
where we have defined the term
\be \label{EQ:NAB-K}
\mathcal{K}(\ell,\Omega)\equiv u^t u_\phi=l/(1-\Omega l)\,.
\ee
We now subtract Eq.~(\ref{terms2s}) from Eq.~(\ref{terms1s})
to find the following condition
%
\begin{eqnarray}
\label{br-ak}
\partial_z\left(K(\ell,\Omega)\right)\partial_\chi\Omega
&=&\underbrace{\partial_z\left(\frac{\partial_\chi p}{\omega}\right)-
  \partial_z\left(\frac{\partial_\chi p}{\omega+b^2}\right)
  -\partial_z\left(\frac{b^z\partial_\chi b_z}{\omega+b^2}
  \right)}_{1^{st}-{\rm
    term}} \nonumber \\
		&&+\underbrace{\partial_\chi\left(\frac{\partial_z
    p}{\omega+b^2}\right)-\partial_z\left(\frac{\partial_\chi p}{\omega}\right)+\partial_\chi\left(\frac{b^2
    \partial_z\ln u^t}{\omega+b^2} \right)}_{2^{nd}-{\rm
    term}}\,,
\end{eqnarray}
%
where, for reasons that will be immediately transparent, we have
added and subtracted the term
$\partial_z\left(\frac{\partial_\chi p}{\omega}\right)$, and where we have used $\partial_z\Omega=0$.
A generalized von Zeipel theorem can therefore be
formulated by saying that $\Omega = {\rm const}$ surfaces
coincide with  $l= {\rm const} $ surfaces
if the right hand side of equation (\ref{br-ak})
vanishes. In fact, when  this is the case,  the left hand side must
satisfy
\be\label{Eq:conditionK}
\partial_z\left(K(\ell,\Omega)\right)\partial_\chi\Omega=0 \Longrightarrow
\partial_z\left(K(\ell,\Omega)\right)=\partial_l\mathcal{K}\partial_z\ell+\partial_\Omega \mathcal{K}\partial_z\Omega=\partial_l\mathcal{K}\partial_z\ell=0\,,
\ee
from which it follows that
$\partial_z l=0$. Since $\partial_z \Omega=0$,
this means
that the surfaces at $\ell={\rm const}$
are the same as those at $\Omega={\rm const}$.
We therefore concentrate our attention on the right side
of Eq.\il(\ref{br-ak}). Under the assumption of barotropic equation of state, namely that
$p=p(e)$,  and using Eq.\il(\ref{Eq:rev-Eq}) we obtain
\be\label{Eq:holp}
\partial_z\partial_\chi\ln u^t=\partial_\chi\partial_z\ln u^t=\partial_\chi\left(\frac{\partial_zp}{\omega}\right)=\partial_z\left(\frac{\partial_\chi p}{\omega}\right)\,.
\ee
As a result, the  ${2^{nd}-{\rm term}}$ of
Eq.\il(\ref{br-ak}) is identically zero, as dictated by Eq.\il(\ref{terms2s}).
On the other hand,
the   ${1^{st}-{\rm term}}$ of Eq.\il(\ref{br-ak}) can be
analyzed to obtain
an expression for the partial derivative  of  $b$ along
$\chi$. In fact, by imposing the vanishing of the
${1^{st}-{\rm term}}$ we find
\be
\label{Eq:d2w}
\quad \partial_z\left[\frac{b^2}{\omega (\omega+b^2)}\partial_\chi p-\frac{b^z \partial_\chi b_z}{\omega +b^2}\right]=0\,.
\ee
Hence, the quantity in the square brackets must be
a function of the coordinate $\chi$ only, and therefore of $\Psi$ only.
Hence, (see Appendix B for the details)
\be
\label{eq:PPsi}
\frac{b^2}{2 (\omega+b^2)}\left[{2}\frac{\partial_\chi p}{\omega}-\partial_\chi\ln (b^2 g_{zz})\right]={\cal P}(\Psi)\,.
\ee
The constraint expressed by Eq.~(\ref{Eq:d2w}) can be further manipulated, leading to
\be
\label{Eq:dfeb}
\partial_z \ln\left(\frac{\omega}{b^2}\right)\left[\partial_\chi\ln (b^2 g_{zz})-2\frac{\partial_\chi p}{\omega}\right]=2\left(\frac{\omega+b^2}{\omega}\right)\partial_\chi\partial_z \ln (b^2\sqrt{-g})\,.
\ee
We emphasize that Eq.~({\ref{Eq:dfeb}) and  Eq.\il(\ref{Eq:for-bz})
form a system of partial differential equations for the magnetic field $b$ and the rest mass density $\rho$.
To recap, when the magnetic field is poloidal,
$\Omega={\rm const}$ surfaces coincide with $\ell={\rm const}$ surfaces if
\begin{enumerate}
\item the equation of state is barotropic, i.e. $p=p(e)$, and
\item
the magnetic field obeys the PDE (\ref{Eq:dfeb}), or, equivalently, if the left hand side of (\ref{eq:PPsi}) is a function of $\Psi$ only,
where {$(z,\chi)$ are the B-O coordinates
such that $b^\chi=0$}.
\end{enumerate}
A   special case for which Eq.\il(\ref{Eq:for-bz}) admits a relatively simple solutions is discussed in Appendix C.
It is also worth
commenting about the relevance of our result in the context of the solution of the Grad-Shafranov (GS) equation, which is a highly non-linear partial differential equation  in the unknown flux function $\Psi$.
Notoriously, the  GS equation is used to compute stationary and axisymmetric
magnetohydrodynamics solutions admitting a number of integral  of motions, which depend only on $\Psi$.
In the relativistic regime it has been analyzed by several authors in various
physical contexts, including astrophysical jets, relativistic stars and magnetospheres of compact objects
(see \cite{Camenzind1987}, \cite{Nitta1991}, \cite{Fendt1996}, \cite{Ioka2003}, \cite{Beskin:2009}, \cite{Markakis2011}).
The GS equation, however, provides no information about the relation among $\Omega$ and $\ell$ and the solution that is obtained in general does not satisfy
the conditions of von Zeipel theorem. Our results show that, in order for the GS equation
to give a solution with the property that $\Omega=\Omega(\ell)$, there must by an additional function ${\cal P}(\Psi)$, defined by (\ref{eq:PPsi}) in B-O coordinates, which depends only on $\Psi$.

\section{Conclusions}
We have studied the conditions under which the surfaces
of constant angular velocity $\Omega$ coincide with those
of constant specific angular momentum $\ell$ for a
stationary and axisymmetric magnetized
perfect fluid in circular motion around a compact object.
In the case of a purely toroidal magnetic field,
such conditions amount to  integrability conditions,
both in the Newtonian and in the relativistic regime.
The relativistic treatment of the poloidal magnetic field is more involved,
and it is convenient to adopt suitable coordinates
$(z,\chi)$,
such that $\chi$ is
constant along $\Psi={\rm const}$ surfaces,
while $z$ is
constant along the normals to those surfaces.
In this way it is possible to show that
$\Omega={\rm const}$ surfaces coincide with $\ell={\rm const}$ surfaces if
the equation of state is barotropic, and if
a function ${\cal P}(\Psi)$ exists, given by Eq.~(\ref{eq:PPsi}), which depends only on $\Psi$.
These results become relevant when the construction of an equilibrium solution with a poloidal magnetic field is considered, a task that we will consider in the future through the numerical solution of the Grad-Shafranov equation.

\begin{acknowledgements}
We thank Marek Abramowicz for useful discussions
undertaken at the beginning of this project. D.P. would like to thank the institutional support
of   the Faculty of Philosophy and Science of the Silesian University of Opava.
\end{acknowledgements}

\bibliographystyle{spmpsci}      
\bibliography{references}

\appendix
\section{Example of B-O coordinates in the Schwarzschild metric}
\label{AppendixC}

As an example of how to choose the B-O coordinates $(z,\chi)$ in
a specific context, let us  briefly consider the case of
a magnetized fluid rotating  with constant specific angular momentum
$\ell$ in a Schwarzschild metric.
From the iso-rotation law we know that
$\Omega=\Omega(\Psi)$, and since we are looking for a
coordinate $\chi=\chi(\Psi)$, we can simply take
$\chi=\Omega$. However, in the Schwarzschild metric
$\Omega=-g_{tt}\ell/g_{\phi\phi}$ and, since $\ell={\rm const}$, we can choose
\begin{equation}
\chi = \frac{(1-2/r)}{r^2\sin^2\theta}\,.
\end{equation}
On the other hand, the coordinate $z$ can be computed
from the requirement that the orthogonality between  $z$
and $\chi$ is preserved,
i.e. $g_{z\chi}=0$. Straightforward metric coefficients
transformations yield
\begin{equation}
z=(r-3)\cos\theta \,.
\end{equation}
In terms of such B-O coordinates it is then possible to write the constraints
expressed by Eq.~(\ref{Eq:for-bz}) and by
Eq.~(\ref{Eq:dfeb}) to guarantee that $\Omega=\Omega(\ell)$.


%
\section{Derivation of Eq.~(\ref{Eq:dfeb})}
\label{appendixB}
We first note that, in the coordinates $(\chi,z)$, $b^2=b_{\alpha}b^{\alpha}=b_z b^z$, and
\bea
\label{Eq:i}
b^z&=& \frac{b}{\sqrt{g_{zz}}}\,,\quad b_z= b \sqrt{g_{zz}}\,,\\
\label{Eq:adeb}
b^z\partial_i b_z&=&  \frac{\partial_i b^2}{2}-\frac{1}{2}\mathcal{M}_ib^2,\quad \mathcal{M}_i \equiv g_{zz}\partial_i g^{zz}=\partial_i\ln g^{zz}=-\partial_i\ln g_{zz}\,.
\eea
Using these definitions, it is possible to rewrite
Eq.\il(\ref{Eq:d2w}) as
\be\label{Eq:d2wc}
\partial_z\left\{\frac{b^2}{2 (\omega+b^2)}\left[{2}\frac{\partial_\chi p}{\omega}-\partial_\chi\ln b^2-\partial_\chi\ln g_{zz}\right]\right\}=0.
\ee
If we expand the derivatives in Eq.~(\ref{Eq:d2wc}), we obtain
 \be
 2 \partial_z\left(\frac{\partial_\chi p}{\omega}\right)-\partial_z\partial_\chi\ln(b^2 g_{zz})+\partial_z\ln\left(\frac{b^2}{\omega+b^2}\right) \left[2\frac{\partial_\chi p}{\omega}-\partial_\chi \ln(b^2 g_{zz})\right]=0\,,
 \ee
which, by using Eq.\il(\ref{Eq:for-bz}), can also be written as
\be\label{tobesim}
-2\partial_\chi\partial_z \ln(b^2\sqrt{-g})+\partial_z\ln\left(\frac{b^2}{\omega+b^2}\right) \left[2\frac{\partial_\chi p}{\omega}-\partial_\chi \ln(b^2 g_{zz})\right]=0\,,
\ee
%
or, equivalently
\be\label{Eq:dfe}
\partial_\chi\ln (b^2 g_{zz})=2\frac{\partial_\chi p}{\omega}-\frac{2\partial_\chi\partial_z \ln ((b^2\sqrt{-g}))}{\partial_z \ln\left(\frac{b^2}{\omega+b^2}\right)}\,.
\ee
The term $\partial_z\ln\left(\frac{b^2}{\omega +b^2}\right)$ can now be replaced using the identity
\be
\partial_z\ln (\omega+b^2)=\partial_z\ln b^2+\frac{\omega}{\omega+b^2}\partial_z\ln\left(\frac{\omega}{b^2}\right)\,,
\ee
thus allowing to rewrite
Eq.\il(\ref{Eq:dfe}) as
\be\label{Eq:rew}
\partial_z \ln\left(\frac{\omega}{b^2}\right)\left[\partial_\chi\ln (b^2 g_{zz})-2\frac{\partial_\chi p}{\omega}\right]=2\left(\frac{\omega+b^2}{\omega}\right)\partial_\chi\partial_z \ln (b^2\sqrt{-g})\,,
\ee
which is Eq.~(\ref{Eq:dfeb}) of the main text. An alternative expression in which a single term with a first order derivative of the magnetic field is also possible and it is given by
\be\label{Eq:dfeb-b}
\partial_\chi\ln b= -\left(\frac{b^2+\omega}{2\omega}\right)\frac{\partial_\chi\left(\frac{2\partial_zp}{\omega}-\partial_z\ln \frac{g_{zz}}{\sqrt{-g}}\right)}{ \partial_z\ln{\sqrt{\frac{-g\omega}{g_{zz}}}}+\frac{\partial_zp}{\omega}}+\frac{\partial_\chi p}{\omega}-\partial_\chi\ln \sqrt{g_{zz}}\,.
\ee
%


%
\section{Analysis of a special case: the polytropic equation of state}
\label{AppendixC}

A particular case  for which Eq.\il(\ref{Eq:for-bz}) allows
a simple integration is for  a polytropic equations of state of the type
$p=k \omega^{\Gamma}$, with $ \Gamma $  and $k$  constants. In such circumstances  a solution of Eq.\il(\ref{Eq:for-bz}) is
\bea
\label{Eq:particular1}
b=f(\chi)e^{-\frac{k \Gamma   \omega^{\Gamma-1}}{\Gamma-1}}\frac{ \sqrt{g_{zz}} }{\sqrt{-g}},\quad \Gamma\neq 1 \,,
\eea
where $f(\chi)$ is an arbitrary function of $\chi$, to be fixed through
Eq.~(\ref{Eq:dfeb}), or, alternatively, through Eq.\il(\ref{Eq:dfeb-b}).
Hence, Eq.\il(\ref{Eq:particular1})  expresses the general form of the magnetic field, written in B-O coordinates, when
a polytropic equation of state is adopted. The fulfillment of von Zeipel's hypothesis through
the additional condition (\ref{Eq:dfeb-b}) provides a  constrain on the function $ f(\chi) $, which must satisfy
\be
\label{Eq:foromega}
\frac{\partial_\chi f}{f}+f^2 f_1(\chi, z)+f_2(\chi, z)=0,
\ee
where $f_1$ and $f_2$ are two functions of $(\chi, z)$ through  $\omega$, the metric  $\mathbf{g}$ and
their derivatives.   As $f$ is a function of $\chi$ only,   after taking the $z$ derivative of Eq.~(\ref{Eq:foromega}),
we find the relation $f(\chi)^2=-\partial_z f_2/\partial_z f_1$, which could be in principle used to determine  $\omega$, and ultimately the rest mass density $\rho$, in a given spacetime metric. Although the explicit form of $f_1$ and $f_2$ can be extracted from Eq.~(\ref{Eq:dfeb-b}), the relevant point is that, at least  formally, the  general solution of Eq.~(\ref{Eq:foromega}) is
\be
f(\chi)=\mp \frac{e^{\int_1^\chi -f_2(t,z) \, dt}}{\sqrt{c_1+2 \int _1^\chi e^{2 \int_1^{s} -f_2(t,z) \, dt} f_1(s,z)ds}}\,,
\ee
where $c_1$ is a constant.}

\end{document}